# Disturbed Fossil Group Galaxy NGC 1132

Dong-Woo Kim, Craig Anderson, Doug Burke, Giuseppina Fabbiano,
Antonella Fruscione, Jen Lauer, Michael McCollough, Doug Morgan,
Amy Mossman, Ewan O'Sullivan, Alessandro Paggi, Saeqa Vrtilek

Smithsonian Astrophysical Observatory
60 Garden Street, Cambridge, MA 02138, USA

Ginevra Trinchieri

INAF-Osservatorio Astronomico di Brera, Milan Italy

(Dec. 11, 2017)

**abstract**

We have analyzed the Chandra archival data of NGC 1132, a well-known fossil group, i.e. a system expected to be old and relaxed long after the giant elliptical galaxy assembly. Instead, the Chandra data reveal that the hot gas morphology is disturbed and asymmetrical, with a cold front following a possible bow shock. We discuss possible origins of the disturbed hot halo, including sloshing by a nearby object, merger, ram pressure by external hotter gas and nuclear outburst. We consider that the first two mechanisms are likely explanations for the disturbed hot halo, with a slight preference for a minor merger with a low impact parameter because of the match with simulations and previous optical observations. In this case, NGC 1132 may be a rare example of unusual late mergers seen in recent simulations. Regardless of the origin of the disturbed hot halo, the paradigm of the fossil system needs to be reconsidered.

# 1. INTRODUCTION

Isolated giant elliptical galaxies associated with large, extended hot halos are thought to be the end product of past galaxy mergers within a normal group, i.e. 'Fossil Groups' (see Ponman et al. 1994). The fossil group is commonly defined as an X-ray luminous ($L_X > 2 \times 10^{42}$ erg s$^{-1}$) group where a single elliptical galaxy dominates the optical luminosity of the entire group with a large magnitude gap ($\Delta m_{12} > 2$ mag in R-band) between 1$^{st}$ and 2$^{nd}$ brightest galaxies within a half of the virial radius (Jones et al. 2003). This type of system was also called an X-ray over-luminous elliptical galaxy (OLEG) by Vikhlinin et al. (1999) and an isolated X-ray over-luminous elliptical galaxy (IOLEG) by Yoshioka et al. (2004). An alternative definition (e.g., $\Delta m_{14} > 2.5$ mag between the 1$^{st}$ and 4$^{th}$ brightest galaxy) is sometimes used (e.g., Dariush et al. 2010).

Recent studies of local fossil groups and clusters show that their X-ray luminosities are in the range $10^{42} - 10^{45}$ erg s$^{-1}$ (Jones et al. 2003; Girardi et al. 2014, Bharadwaj et al. 2016), 1-3 orders of magnitude higher than those of typical giant elliptical galaxies with a similar optical luminosity (e.g., Kim & Fabbiano 2015). These large X-ray luminosities indicate the presence of substantial dark matter halos. Their X-ray surface brightness is usually extended and smooth, as expected for an old relaxed system (e.g., Eigenthaler & Zeilinger 2013, Bharadwaj et al. 2016). The $\Delta m_{12}$ (or $\Delta m_{14}$) condition further ensures that the system is indeed evolved long after the galaxy assembly by mergers. These properties, and the generally old stellar populations (with ages > 3 Gyr) suggested galaxies at the end of their evolution (e.g., Jones et al. 2000; Khosroshahi et al. 2006).

Recent simulations, however, suggest an alternative scenario. The growth of the massive galaxy may continue even after the group has entered the fossil phase by renewed infall from its environment (von Benda-Beckmann et al. 2008; Díaz-Giménez et al. 2008; Dariush et al. 2010; Kanagusuku et al. 2015). For some fossil groups, the last major merger might have occurred as recently as < 2 Gyr ago (von Benda-Beckmann et al. 2008). These numerical simulations indicate that signs of recent merging should be fairly common in first-ranked fossil group galaxies, and the paradigm of fossil groups as relaxed, undisturbed systems needs to be reconsidered (Díaz-Giménez et al. 2008, Alamo-Martinez 2012).

NGC 1132 has been considered as a fossil group by a number of previous studies (e.g., Mulchaey & Zabludoff 1999; Yoshioka et al. 2004, Eigenthaler & Zeilinger 2013; Lovisari et al. 2015, Bharadwaj et al. 2016), even called as a prototype (Alamo-Martínez et al. 2012). Although a nearby spiral galaxy (NGC 1126) is 2.1 mag fainter in B but only 1.6 mag fainter in K as pointed out by Sun et al. (2009), NGC 1132 is similar to other fossil groups (see also Eigenthaler & Zeilinger 2013). The previous ASCA and XMM-Newton X-ray observations of NGC 1132 show an extended X-ray luminous halo with $L_X$ = several $\times 10^{42}$ erg s$^{-1}$, consistent with those of fossil groups (Mulchaey & Zabludoff 1999; Yoshioka et al. 2004; Lovisari et al. 2015, Bharadwaj et al. 2016). NGC 1132 is also known to be a slow rotator (Veale et al. 2017) with a core radial profile at the center (Alamo-Martinez 2012), again similar to typical old giant elliptical galaxies.

Analyzing the high-resolution Chandra archival data of NGC 1132 we find that the hot halo is far from relaxed, suggesting that it may provide observational evidence of a perturbed fossil group. In this paper, we present the results of our analysis of the disturbed hot halo and discuss possible origins for this unusual fossil system. In section 2, we describe the Chandra observations, our analysis techniques and statistical significance of our findings. In section 3, we discuss various possibilities of the origin of the disturbed hot gas and we summarize our results in section 4. We adopt D=95 Mpc, taken from NED[1]. At this distance, 1 arcmin corresponds to 27.6 kpc. Throughout this paper, we quote an error in 1σ significance level.

## 2. CHANDRA OBSERVATIONS

NGC 1132 was observed twice, in 1999 and 2003, both with ACIS-S (see Table 1). The total effective exposure time is 38.3 ksec, after excluding about 30% due to background flares. We analyzed the Chandra data as a part of the Chandra Galaxy Atlas project which is described in detail by Kim et al. (in prep). Here we describe briefly key steps that were applied:

Table 1. Chandra Observation log

| obsid | Obs date | Instrument | Net exp (ksec) | RA | DEC |
|---|---|---|---|---|---|
| 801 | 1999 Dec. 10 | ACIS-S | 11.7 | 02:52:51.6 | -01:16:32.8 |
| 3576 | 2003 Nov. 16 | ACIS-S | 26.6 | 02:52:52.0 | -01:16:29.6 |

First, we exclude all point sources detected by a CIAO[2] tool, **_wavdetect_**. The point sources are mostly low-mass X-ray binaries (LMXB) in NGC 1132 as well as background galaxies and AGNs (e.g., Green et al. 2004). To determine the size of each point source, we make a PSF for each source for each observation, using MARX[3]. The point source elliptical region is filled with values interpolated from surrounding pixels by using a series of CIAO tools, **_roi_**, **_splitroi_** and **_dmfilth_**. We generate a point-source-excluded, filled, exposure-corrected image using the CIAO tool, **_fluximage_**. Then we smooth the diffuse image with two Gaussian kernels: a smaller scale kernel with a Gaussian σ =3.5 arcsec, and a larger scale kernel with σ = 10 arcsec. The resulting images shown in Figure 1 represent the hot gas in the 0.5-2keV energy band. The small-scale smoothing (Figure 1a) shows a sharp discontinuity east of the galaxy center (the edges are marked by blue lines in Figure 2) and a hint of asymmetric emission toward the west. The large-scale smoothing (Figure 1b) shows the extent of the asymmetry toward the west. The $D_{25}$ ellipse (semi-major axis = 1.3 arcmin or 35 kpc at D=95 Mpc) is marked in both figures. The sharp discontinuity and extended hot gas form a head-tail structure, suggesting that the hot gas is moving toward the east or exposed to an external pressure in the eastern side, pushing towards the west.

While the images in Figure 1 are by themselves quite intriguing, the 2D temperature maps shown in Figure 2 suggests even more interesting features. The temperature maps were produced by applying the relevant binning methods from the Chandra Galaxy Atlas (CGA) project. In CGA, we use four adaptive binning methods to characterize the 2D spectral properties: (a) annulus binning with adaptively determined inner and outer radii (b) weighted Voronoi tessellation (WVT) adaptive binning (Diehl & Statler 2006), (c) contour binning which is similar to WVT but further takes into account the fact that similar surface brightness regions have similar spectral properties (Sanders 2006), and (d) hybrid binning which maintains a high S/N by extracting spectra from larger circular regions while keeping the spatial resolution in finer spatial grids (O'Sullivan et al. 2014). The fourth method uses neighboring bins which are not independent (hence statistics are not straightforward), but provides complementary information at higher spatial resolution, which may be lost in the first three methods. Methods (a) and (b) are not appropriate for these data, because the hot gas emission is not azimuthally symmetric, and data have limited statistics. Once spatial adaptive binning is done, the X-ray spectra are extracted from each spatial bin. The spectral extraction is done per observation per chip. The corresponding arf and rmf files are also extracted per observation per chip to take into account time- and position-dependent ACIS responses. For background emission, we download the blank sky data from the Chandra archive, re-project them to the same tangent plane as each observation and rescale them to match the rate at higher energies (9-12 keV) where the photons are primarily from the background (Markevitch 2003). To confirm the validity of the sky background and to check temporal and spatial variations of the soft X-ray background, we used the off-axis, source-free region from the same observations and did not find any significant difference.

---

[1] http://ned.ipac.caltech.edu
[2] http://cxc.harvard.edu/ciao/
[3] http://space.mit.edu/cxc/marx/

In Figure 2 we show the temperature map obtained with the last two binning methods (S/N=20 in each spatial bin). We found no significant difference in the resulting temperatures derived using one or two component models (APEC for hot gas or APEC + power-law to account for undetected point sources). We fixed the abundance at solar, but the temperature is not sensitive to the abundance, although the normalization (hence emissivity and density) is. The temperature maps show that the gas inside the discontinuity seen in Figure 1a (the edges of which are represented by the two blue lines in Figure 2) is cooler (kT = 0.8 - 1 keV) than the surrounding gas (kT ~ 1.2 keV), suggesting a cold front. The gas ahead of the cold front (near the $D_{25}$ ellipse of NGC 1132) is hotter (kT = 1.25 - 1.35 keV) than the gas in the outer halo (kT ~ 1 keV), possibly indicating that there is a shock propagating ahead of the cold front.

```
              Table 2. Density profile - best fit parameters and errors
-------------------------------------------------------------------------------------
              power-law slope                           density jump
         r < 10       r=10-30      r > 30        at 10 kpc      at 30 kpc      chi2      dof
-------------------------------------------------------------------------------------
Case A   1.24 (0.02)  0.97 (0.02)    --          1.48 (0.12)       --          14.7      14
Case B   1.24 (0.02)  0.85 (0.06)  0.76 (0.08)   1.51 (0.33)    1.35 (0.84)    12.2      12
-------------------------------------------------------------------------------------
Case A: fit with a two power-law density profile with 1 density jump.
Case B: fit with a three power-law density profile with 2 density jumps.
```

To further quantify the discontinuity and the related jump in gas properties, we selected a pie sector spanning PA=30-120°. The surface brightness and temperature profiles in this pie sector are shown in Figure 3 and Figure 4, respectively. The surface brightness (SB) profile was made in the radial bins with the width increasing with increasing r (from 1 to 5 kpc). The radial profile follows a power law at the outskirts (r=30-60 kpc; blue dotted line). However, the SB profile in the inner region deviates strongly from the extrapolation of this power law and suggests the presence of an edge at r ~ 10 kpc (blue bar) and a possible $2^{nd}$ edge at r ~ 30 kpc

(red bar). The first one corresponds to the cold front seen in Figure 1 and the second one to the possible shock front seen in Figure 2. To estimate the density discontinuities at the edges, we applied (Case A) a two power-law density profile with one jump and (Case B) a three power-law density profile with two jumps. Integrating the emissivity along the line of sight and fitting the 2D projected profile, we determine the best fit parameters of the density profiles with corresponding errors and list them in Table 2. At the $1^{st}$ edge (r=10 kpc), the density jumps by a factor of 1.5 (± 0.12 in Case A; ± 0.33 in Case B). At the possible $2^{nd}$ edge, the density jumps by 1.3, but with a large error. Based on the F distribution (Bevington 1969), the statistical improvement with an additional $2^{nd}$ jump is marginal with a probability of 0.11 for exceeding the given statistic.

The temperature profile in the same pie sector is shown in Figure 4. The bins are adaptively determined such that each bin has at least S/N=20 to ensure the reliable temperature measurement. The typical temperature error is a few percent (0.03-0.09 keV) and the reduce $\chi^2$ is always close to 1 or less. As seen in Figure 2, the gas at r=15-30 kpc is nominally hotter (1.34 ± 0.05 keV) than the gas (1.08 ± 0.06 keV) inside and outside this region. The statistical significance of the temperature jump is 2.9σ (see Table 3 for the jump conditions and Section 3.2 for further discussions). In Figure 5, to illustrate the data quality and goodness of the fit, we compare the spectral fitting results from three radial bins (the hotter, possibly shocked gas and inside/outside the hotter gas). As the gas temperature is mainly determined by the peak energy (Fe L complex), the energy differences of these spectral peaks indicate that the temperature can be reliably measured and that the background and the hard component are not important. For comparison, we also extract spectra in the opposite side (PA = 224-315°). In this complementary wedge (cyan points in Figure 4) there is no significant temperature jump. We note that the 3D temperature could reveal the temperature jump more significantly (see below for our approximate measurements when the multi-temperature gas is present). Deeper observations will be needed to solidly confirm the presence of the shock. If confirmed, NGC 1132 may yield the first detection of a galaxy-scale bow shock propagating ahead of the cold front in a ~1 keV hot plasma.

Table 3. Jump conditions

|  |  | density jump | T jump | pressure jump | entropy jump |
|---|---|---|---|---|---|
| Case A | cold front (r=10) | 1.48 (0.12) | 0.80 (0.05) | 1.18 (0.12) | 0.61 (0.05) |
| Case B | cold front (r=10) | 1.51 (0.33) | 0.80 (0.05) | 1.21 (0.28) | 0.61 (0.10) |
| Case B | shock front (r=30) | 1.35 (0.84) | 1.23 (0.08) | 1.66 (1.04) | 1.06 (0.42) |

## 3. DISCUSSION

In contrast to the previous expectation of a prototype fossil system, the hot gas morphology of NGC 1132 indicates that this system is quite disturbed with an asymmetric hot halo, including a sharp edge to the east (and a possible shock), and extended emission to the W. These results do not conform with the old, relaxed fossil group scenario, and suggest a rare case of a perturbed fossil group, as predicted by recent simulations (e.g., von Benda- Beckmann et al. 2008; Díaz-

Giménez et al. 2008; Dariush et al. 2010; Kanagusuku et al. 2016). We discuss the possible origin of this disturbed fossil system below.

### 3.1 Interaction with environment

**A. Sloshing by a nearby galaxy**

Chandra observations have revealed cold fronts in many gas-rich elliptical galaxies (as well as in groups and clusters). The cold front can be developed by 'sloshing' of the ISM of the dominant galaxy, perturbed by the passage of a nearby large companion galaxy (Ascasibar & Markevitch 2006). Well known examples are NGC 507 (with NGC 499, 5′ away or 90 kpc in projection, with $\Delta m_B = 0.7$ mag or $\Delta m_K = 0.4$ mag), NGC 1399 (with NGC 1404, 10′ away or 60 kpc in projection, with $\Delta m_B = \Delta m_K = 0.5$ mag), NGC 5044 (with NGC 5037, 14′ away or 130 kpc in projection, with $\Delta m_B = \Delta m_K = 0.9$ mag; and with NGC 5054, 27′ away or 250 kpc in projection with $\Delta m_B = -0.5$ mag or $\Delta m_K = -0.1$ mag, NGC 5054 is brighter than NGC 5044), NGC 5846 (with NGC 5850, 10′ away or 70 kpc in projection, with $\Delta m_B = 0.5$ mag or $\Delta m_K = 1.2$ mag) and NGC 7618 (with UGC 12491= PGC 71014, 14′ away or 300 kpc in projection, with $\Delta m_B = 0.9$ mag or $\Delta m_K = 0.8$ mag).

The hot halos of these systems have temperatures (~1 keV) similar to that measured in NGC 1132, suggesting that their total masses are similar. However, they have companions within a few x 100 kpc (or roughly one third of $r_{200}$ for 1 keV systems) with comparable stellar luminosity (with $\Delta m_B < 0.9$ mag or $\Delta m_K < 1.2$ mag), or a luminosity ratio of 2:1 or 3:1. NGC 1132, instead, does not have a nearby companion which is large enough to initiate sloshing. A few nearby (within 5' in projection) galaxies are too small, 100 times fainter than NGC 1132 in their B-band optical luminosities. A spiral galaxy, NGC 1126, 7 (or 5) times fainter in B (or in K) than NGC 1132 is 8.4' west (or 230 kpc in projection) from NGC 1132. We note that the total mass ratio will be larger than the luminosity ratio because the main galaxy (NGC 1132) is sitting at the bottom of the group potential. Although NGC 1126 may have its own group halo, it is unlikely that the spiral galaxy is the BCG of a small group which hosts a significant amount of dark matter. The real separation can be as large as several Mpc behind NGC 1132, based on their velocity difference ($\Delta v = 440$ km s$^{-1}$). In this case, NGC 1126 would be beyond the virial radius and would have no effect. If NGC 1126 is near NGC 1132, it will move away at the opposite side (west) of the edges of NGC 1132 with an impact parameter of a few x 100 kpc. Given the large mass ratio and large impact parameter, the collision would be minor. Another bright galaxy, 5 times fainter than NGC 1132 ($\Delta m_B = 1.8$ mag, $\Delta m_K = 1.5$ mag), PGC 10856, is 0.5 deg away from NGC 1132. The projected distance alone, puts it at 800 kpc away, which is about twice $r_{500}$ (Sun et al. 2009) and close to the virial radius of a halo with 1 keV (e.g., Sanderson et al. 2003). Therefore, it is unlikely that PGC 10856 could have caused the sloshing in NGC 1132.

Moreover, the intensity and temperature maps of NGC 1132 do not agree with the sloshing simulations with a large mass ratio and a large impact parameter (e.g., ZuHone & Kowalik et al. 2016 – however their simulations are for clusters with larger scales and higher temperatures). However, we cannot rule out the sloshing which might be triggered by an extreme object, e.g., an optically small galaxy with a high mass-to-light ratio, or even dark matter only structures which have lost all their gas (Ascasibar & Markevitch 2006).

A distinct signature of the sloshing is a spiral feature as that seen in NGC 7618 (Roediger et el. 2012) and multiple edges in opposite sides, also confirmed in simulations (e.g., Ascasibar

& Markevitch 2006; Zuhone & Roediger 2016). We do not see such a spiral feature, nor multiple edges in the Chandra image (Figure 1 and 2). To further check this, we produced a residual image after subtracting a best-fit 2D beta model from the exposure corrected, smoothed (a Gaussian sigma= 20 arcsec) image (see Figure 6). The most significant feature is the extended hot gas toward the west, expected because of the asymmetric gas distribution. The hint of a curved feature from the western extension at r = 60-80 kpc from the center toward the north-west direction is too faint to be conclusively interpreted as the spiral feature expected from the sloshing simulations. We have also applied the unsharp masking technique with various pairs of smoothing factors, but found no significant feature possibly associated with sloshing. Deeper observations are needed to see any spiral pattern that could prove the sloshing hypothesis.

In a large (~10') scale, the ASCA GIS image in Figure 1 of Yoshioka et al. (2004) does not show any distinct asymmetrical feature. We have also checked the XMM-Newton images (20 ksec after excluding background flares, downloaded from XSA[4]). The smoothed MOS image reveals that the hot gas is more extended toward the west (as in Figure 1b), but again no spiral feature can be identified.

**B. Recent minor merger with a low impact parameter**

Given the lack of luminous nearby galaxies in the NGC1132 group, one may consider that the perturber might have been already disrupted and perhaps merged into NGC 1132 after several orbital passages in a few Gyr. However, it is unlikely that the rather sharp observed SB feature (at the cold front) can last that long as the disturbance in the hot plasma can last only for about 1 Gyr (e.g., Markevitch & Vikhlinin 2007). Instead, a viable scenario may be that a small galaxy has fallen from the west on a radial orbit and is now passing close to NGC 1132. Interestingly, if the hotter gas found at r = 15-30 kpc east from the center is indeed caused by a shock (see more section 3.2), this hypothesis may explain the observed features by the shock front propagating ahead of the cold front. In this case, the impact parameter is expected to be small, as suggested by the rather symmetric SB roughly in the N-S direction (in contrast to the asymmetric SB in the E-W direction) in Figure 1a. The observed intensity and temperature maps look similar to the first passage of a minor merger with a low impact parameter (e.g., see simulations of the R=1:10 and b=0 case right after the passage through the center by ZuHone & Kowalik et al. 2016). Although no small, secondary galaxy is observed near the position of the front, it should be noted that a small system could be difficult to detect if it falls behind NGC 1132 or has been tidally disrupted by the encounter.

Recent HST optical observations of NGC 1132 (Alamo-Mart'inez et al. 2012) have revealed evidence of galaxy interactions, including dust lanes in the inner regions and shells in the residual image. The optical stellar shells at r < 30 kpc may correspond to a minor merger about ~1 Gyr ago (e.g., Ebrova 2013). This past merger which had caused the remaining shells may not be responsible for the observed hot gas structures which were plausibly formed more recently. Nonetheless, they may indicate multiple minor merges in the last Gyr, suggesting that the merger activity has continued to the recent past. Alamo-Mart'inez et al. (2012) noted that these features contradict previous claims of no sign of merger activity in fossil groups. This supports the suggestion that the process of galaxy merger makes the fossil state a transitory one, with systems that look like fossils potentially forming at any time, as found in recent cosmological simulations (von Benda-Beckmann et al. 2008; Dariush et al. 2010; Kanagusuku et al. 2015).

---

[4] http://nxsa.esac.esa.int/nxsa-web/#home

**C. Ram pressure by intra group medium**

Another explanation for a cold front is the infall of a non-BCG galaxy embedded in the larger, hotter group/cluster gas. During the infall, the ram pressure of the hotter medium produces a sharp discontinuity in one direction and an extended tail in an opposite direction. A well-known example is NGC 1404 which is infalling toward NGC 1399 in the Fornax cluster (e.g., Su et al. 2017a, b). The X-ray image of NGC 1132 may appear to be similar to those in the above examples. However, being a group dominant galaxy, NGC 1132 is centered on the large-scale emission and there are no other large galaxies nearby (as discussed previously), i.e., there is no other plausible BCG. Therefore, NGC 1132 is sitting at the bottom of the group potential and there is no external, hotter medium for NGC 1132 to move against.

Suppose NGC 1132 is falling toward a hypothetical 1 keV group which primarily consists of dark matter with some gas. For a galaxy falling from $r_{500}$ and passing within 10 kpc of the core, a velocity ~1200 km s$^{-1}$ seems likely. For comparison, in the less massive Stephan's Quintet, the interloper galaxy is falling through the group at ~900 km s$^{-1}$ (Moles et al. 1997). This velocity is inconsistent with (significantly greater than) the velocity measured by the jump condition (see Section 3.2), unless the projected distance is significantly large, or the shock estimate is significantly lower.

**3.2 Is there a shock front in NGC1132?**

While Chandra observations have revealed surface brightness discontinuities in many gas-rich elliptical galaxies, all the reported features are cold fronts. No bow shock, with shock-heating of the gas, has been reported in galaxies. Even in clusters, only a handful of systems are known to have bow shocks (e.g., A520 Markevitch et al. 2005; "Bullet Cluster" 1E0657-56 Markevitch et al. 2006; A2034 Owers et al. 2014). The Chandra data of those shock fronts have been usefully applied as an experimental tool to explore various plasma properties, e.g., the front width by comparing with the mean free path and testing electron-ion equilibrium (see Markevitch & Vikhlinin 2007).

As discussed in Section 2, the gas just outside the SB discontinuity (at r = 15-30 kpc west from the center, near the $D_{25}$ ellipse) is hotter (1.34 ± 0.05 keV) than the surrounding gas (1.08 ± 0.06 keV). The temperature jump is significant at 2.9$\sigma$ level, but with no corresponding density jump although the density jump (by ~1.3) is still allowed due to a large error. Taking these results at face value, we can use the jump condition to measure the gas motion. Given the ~1.2 increase in temperature, we estimate the gas velocity to be ~600 km/s (Mach number ~1.2). With a factor of 2 (±1.3) pressure jump between inside the cold front and outside the shock front at the free stream (Markevitch & Vikhlinin 2007) in the same pie sector, we obtain a similar Mach number (M ~ 1) but with a large error (~40%). If this interpretation is valid, the hot gas could be moving at a slight super-sonic speed. Simply dividing the length scale (considering the unknown inclination) by the estimated speed, we estimate that the corresponding age of the fronts is only a fraction of Gyr, indicating these hot gas features are relatively young. If we are indeed seeing a shock, we can rule out the sloshing with a perturbing object outside NGC 1132 for the origin of the disturbed hot gas (as discussed in section 3.1 A).

One way to differentiate between a shock and a cold front is to look for a discontinuity in the pressure (P ~ n T) and entropy (K ~ T $n^{-2/3}$) profile across the edge. The gas pressure is expected to jump at a shock and to be continuous at a cold front, while the entropy varies in an opposite way. Applying the density and temperature jumps measured at the two edges (Table 3), we find the expected trends, i.e., a bigger jump in P at the shock front than at the cold front and a bigger jump in K at the cold front than at the shock front. However, due to the large error in the density jump at the shock front, these jumps are also comparable with each other.

Mazzotta et al. (2004) pointed out that shock fronts are harder to detect than cold fronts in typical Chandra observations, because a single temperature fit to the X-ray spectra from multi-temperature gas always results in a lower temperature than the average (e.g., L-weighted or EM-weighted) value (see also Vikhlinin 2006). In the simple case of two component plasma with two temperatures, $T_{SPEC}$ (the best-fit value of a single component model) can be considerably lower than the high temperature (or an EM-weighted average temperature), even if the EM of the higher T gas is larger than that of the lower T gas. As an example, which may be relevant to our case, fitting the two-temperature (1 keV and 2 keV) gas emission, $T_{SPEC}$ is ~1.3 keV even if the EM of the hotter gas is ~70% (see Figure 3 in Vikhlinin 2006). Therefore, the real temperature of the hotter region (hence the temperature jump) may be underestimated. To test this possibility, we have refit the spectra from the $2^{nd}$ and $3^{rd}$ bins in Figure 4 with a more complex model (two thermal gas components and one power-law component). In both cases, the multi-temperature model does not improve the statistics, as the best-fit two temperatures are almost identical within the error or the $2^{nd}$ component is too weak.

Strong shocks may be accompanied by non-thermal radio emission (e.g., Bullet Cluster Markevitch et al. 2006; A2034 Owers et al. 2014; see an opposite case of the strong shock with no prominent radio relic in A665, Dasadia et al. 2016). We note that the absence of a strong radio relic in NGC 1132 is not contradicting the shock, as it is expected to be low in a low temperature and low Mach number regime (e.g., Skillman et al. 2013).

**3.3 Can a nuclear outburst disturb the hot gas?**

Hydrodynamic simulations (Ciotti et al. 2017; Pellegrini et al. 2012, Eisenreich et al 2017) show that the AGN duty cycle is low (only a few % or less). During the outburst, which lasts for a few tens of Myr, the gas luminosity and temperature jump significantly ($L_X$ by an order of magnitude or more, $T_X$ by a factor of two or more), and then return to the normal range. At the late stage of the outburst, gas with enhanced surface brightness and higher temperature flowing outward may be seen at a radius of ~10 kpc, as we have observed in Figures 3 and 4 (see Figure 8 in Pellegrini et al. 2012 and Figure 4 in Ciotti et al. 2017). In this case, the gas temperature would also peak toward the center. However, there is no such T peak as NGC 1132 has a cool core. Although we cannot rule out the T peak inside r < 1 kpc, the positive T gradient in 2 kpc < r < 20 kpc are different from that predicted by the simulations. Also, the radio emission of NGC 1132 is relatively weak, 5 mJy at 1.4 GHz from both NVSS and FIRST surveys, again implying that the observed edge may not be related to a recent AGN outburst. Dong et al. (2010) reported a cavity at 4kpc south of the center, which we cannot confirm. Even if this cavity is real, it does not coincide with the edge seen in Figure 1.

## 4. SUMMARY

Analyzing 38 ksec Chandra archival data of NGC 1132, a well-known fossil system, we found that the hot gas is disturbed, in contrast to the conventional view that the hot halo in the fossil system would be relaxed and undisturbed. The Chandra data revealed a cold front at r~10 kpc to the east of the center and an extended hot gas region toward the west. There may be a possible shock at r ~30 kpc just ahead of the cold front.

We find that nuclear outburst and external ram pressure are not likely explanations for the NGC 1132 halo features. A nuclear outburst would predict a temperature peak at the center, inconsistent with the observed cool core. Ram pressure stripping is also unlikely because NGC 1132 is sitting at the bottom of the group potential. Sloshing may be the origin of the observed disturbance in the hot halo. Since there is no clear perturber in this fossil system, this can be done only by an extreme object with a large fraction of dark matter. A recent minor merger of a small galaxy with a small impact parameter may cause the disturbance in the hot halo in this fossil system. This hypothesis is consistent with recent simulations of rejuvenated fossil systems and the evidence of recent galaxy interactions provided by optical observations.

Deeper Chandra observations are needed to put our results on a stronger statistical base, especially regarding the existence of a shock front and a possible spiral feature. If confirmed, the former supports the renewed merger hypothesis and rules out the sloshing, while the latter supports the sloshing hypothesis. For a full interpretation of our results, the detailed features need to be compared with customized hydro simulation of a 1keV system.

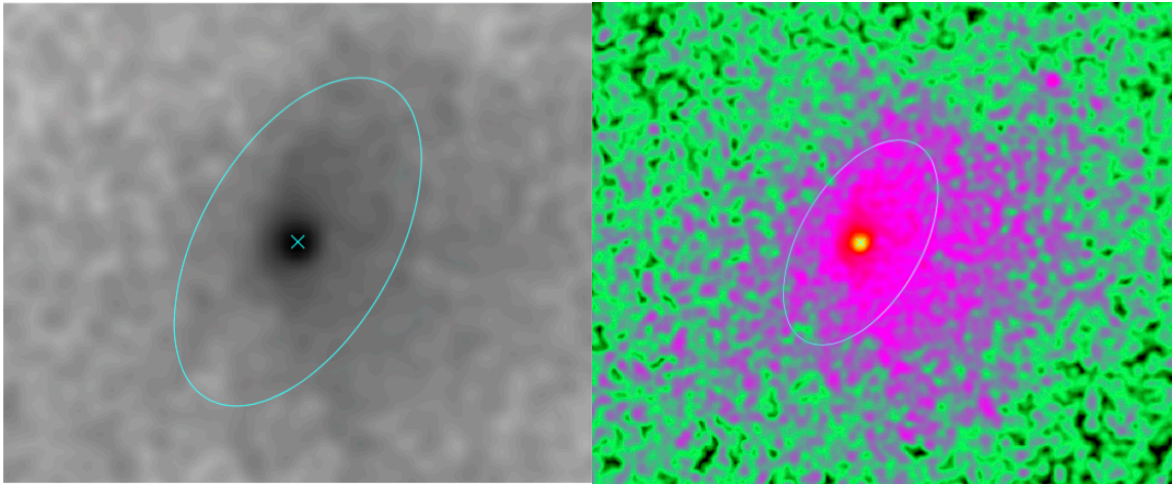

Fig 1. (Left [a]:) Diffuse 0.5-2 keV Chandra emission of NGC 1132 - point sources excluded, filled with neighboring pixel values with Poisson statistics, exposure corrected, smoothed with a Gaussian kernel of sigma = 3.5 arcsec. (Right [b]) same as (Left), but with a Gaussian kernel of sigma = 10 arcsec. The $D_{25}$ ellipse (semi-major axis = 1.3 arcmin or 35 kpc at D=95 Mpc) is shown in both figures.

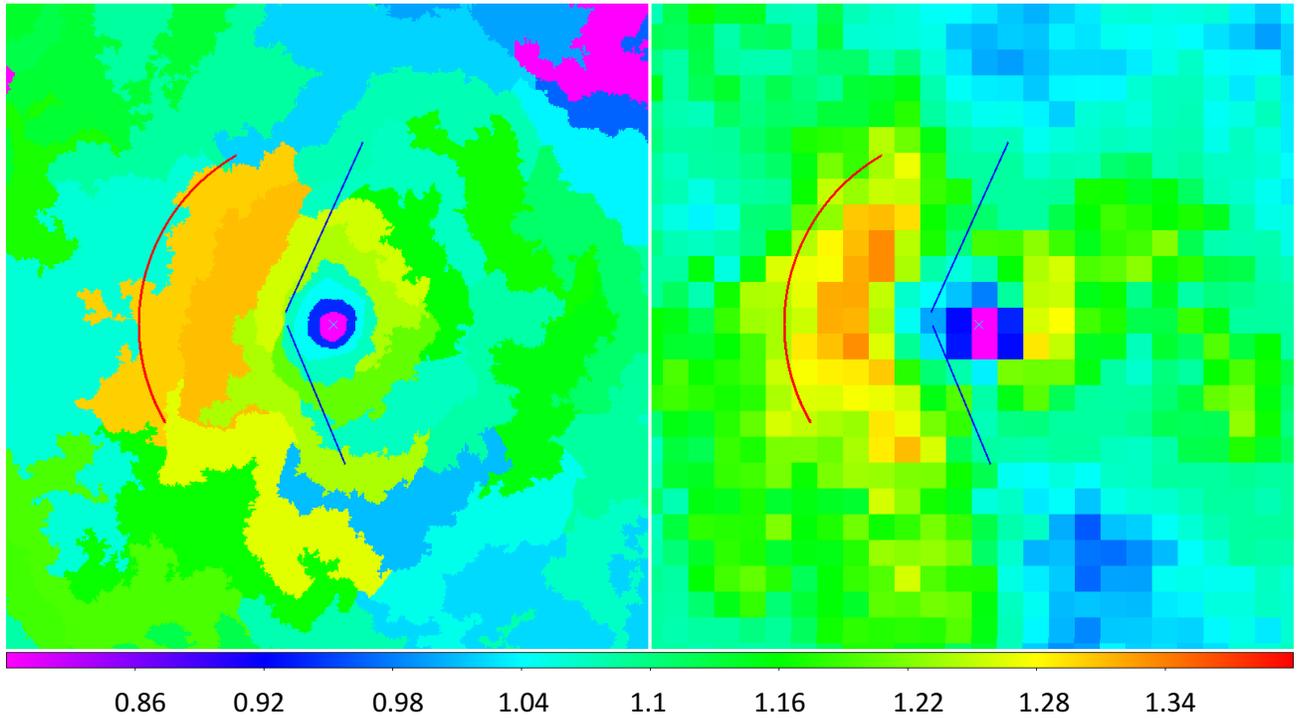

Fig 2. Temperature maps obtained with two methods: (Left [a]:) 2-dimentional adaptive binning by contour binning (Sanders 2006). (Right [b]:) Hybrid binning (O'Sullivan et al 2014). The temperature scale is given at the bottom and ranges from 0.8keV to 1.4 keV. Marked are the location of the discontinuity seen in Fig.1a (in blue) and the possible shock front (in red) at r~30 kpc (the partial circle goes from PA=30°-120°, the pie sector used in Figure 3).

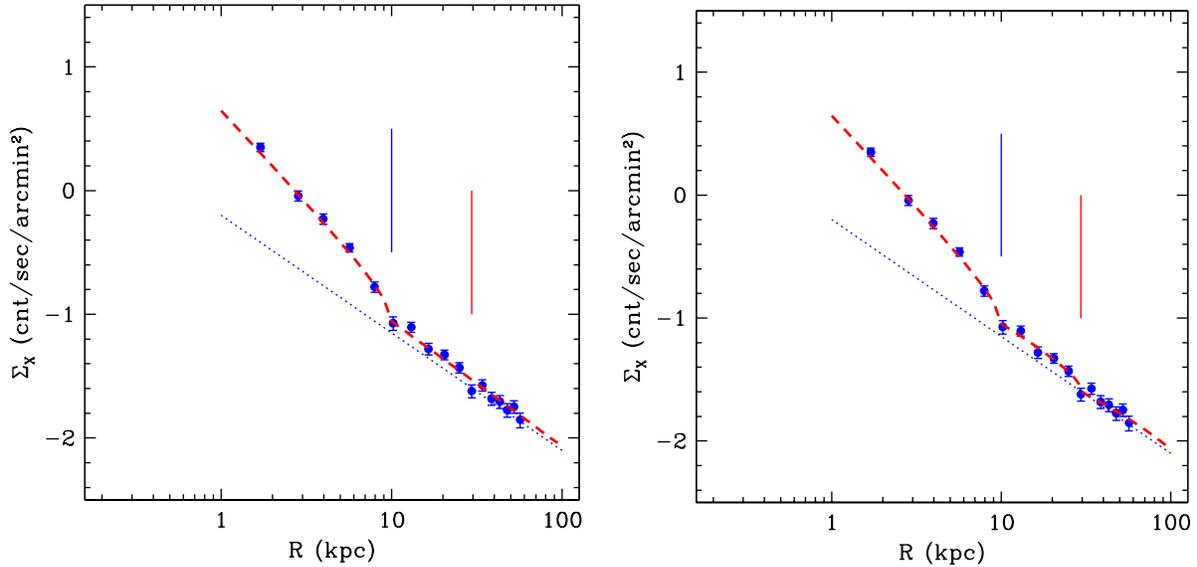

Fig 3. Surface brightness profile in the pie sector (PA=30°-120°) from the center of the galaxy. The blue dotted line is a single power-law determined at the outskirts (r=30-60 kpc) and the red dashed line is obtained from the best-fit 3D density profile. (left) Case A with a two power-law density profile with one jump, (right) Case B with a three power-law density profile with two jumps. The vertical bars mark the two possible edges at ~10 kpc (blue vertical bar) and at ~30 kpc (red bar).

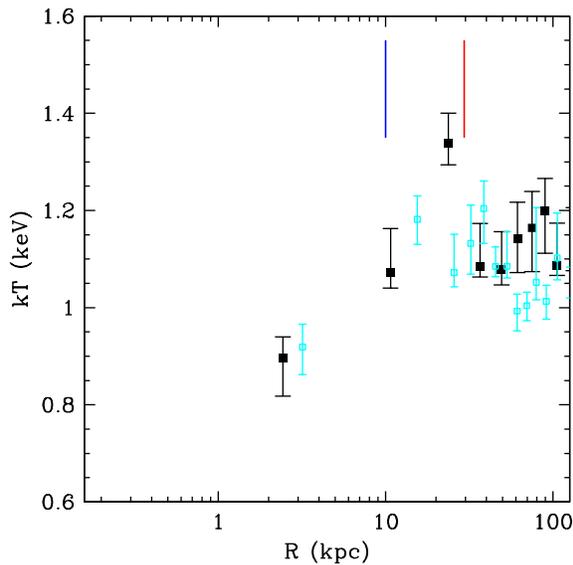

Fig 4. Temperature profiles along the head-side pie sector PA=30°-120° (black filled squares) and the tail-side PA = 224°-315° (cyan open squares). S/N=20 in each spectral bin so that the temperature can be reliably determined with the abundance fixed at solar.

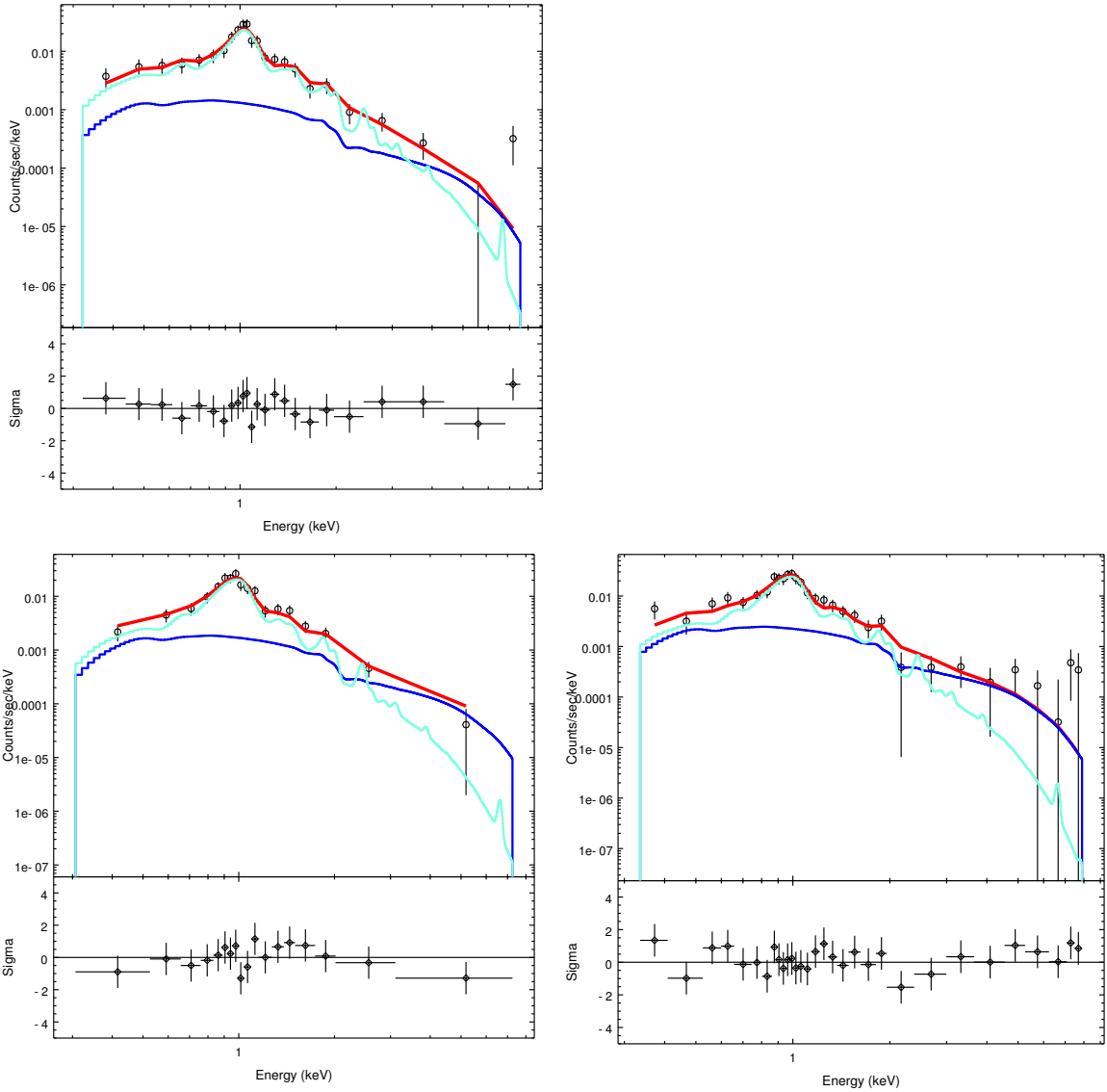

Fig 5. Comparison of spectral fitting results from three radial bins. (a) 1.34 (-0.04, +0.06) keV gas from the possible shock, the 3$^{rd}$ bin in Figure 4, (b) 1.07 (-0.03, +0.09) keV gas from the inner region, the 2$^{nd}$ bin in Figure 4, and (c) 1.08 (-0.03, +0.08) keV gas from the outer region, the 5$^{nd}$ bin in Figure 4. The gas component is in cyan, the hard component (likely from the undetected LMXBs) in blue and the sum in red. In all three cases, the reduced $\chi^2$ is less than 1.

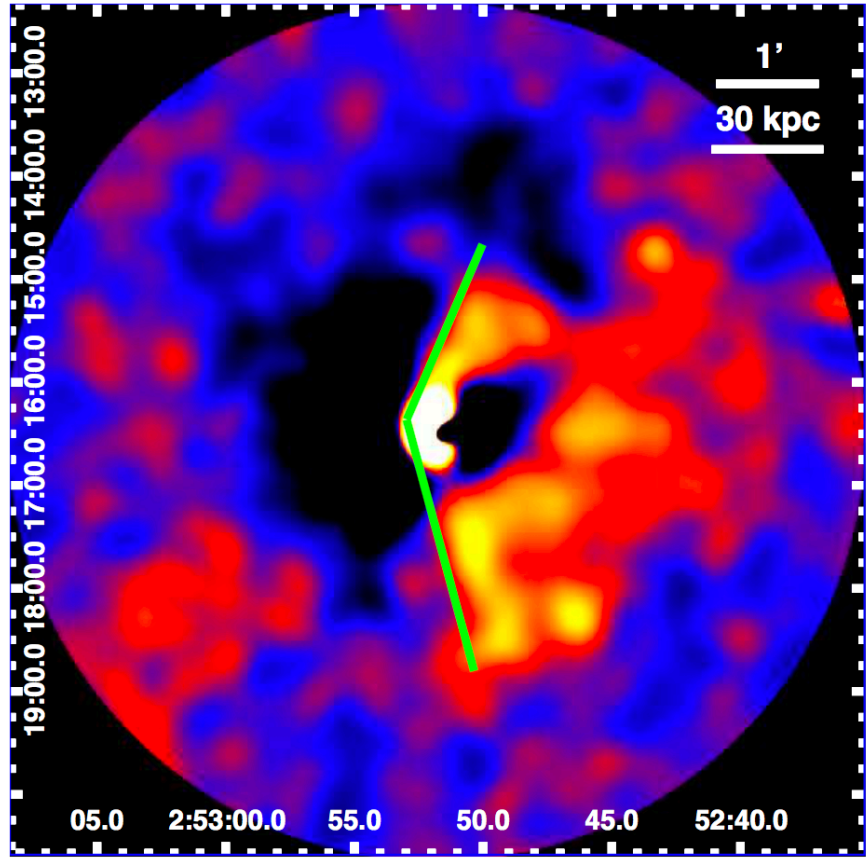

Fig 6. The residual image after subtracting a best-fit 2D beta model from the exposure corrected, smoothed (a Gaussian σ = 20 arcsec) image.